\documentclass[conference]{IEEEtran}
\usepackage{cite}
\usepackage{amsmath,amssymb,amsfonts}
\usepackage{algorithmic}
\usepackage{graphicx}
\usepackage{subfigure}
\usepackage{textcomp}
\usepackage{xcolor}
\def\BibTeX{{\rm B\kern-.05em{\sc i\kern-.025em b}\kern-.08em
		T\kern-.1667em\lower.7ex\hbox{E}\kern-.125emX}}

\usepackage{amsthm}
\usepackage[ruled]{algorithm2e}
\usepackage{mathrsfs}
\usepackage{bm,epsfig,amsthm,url}
\usepackage{indentfirst}
\usepackage{epstopdf}
\usepackage[top=0.75in, left=0.7in, bottom=1.1in, right=0.7in]{geometry}

\theoremstyle{definition}

\begin{document}
	
	\title{Reconfigurable Intelligent Surface for Interference Alignment  in MIMO Device-to-Device Networks}
	
	\author{
		\IEEEauthorblockN{$\text{Min Fu}^{\ast \dagger \S}$, $\text{Yong Zhou}^{\ast}$, and $\text{Yuanming Shi}^{\ast}$
		}\\
		\IEEEauthorblockA{
			$^{\ast}$School of Information Science and Technology, ShanghaiTech University, Shanghai 201210, China \\ 
			$^{\dagger}$Shanghai Institute of Microsystem and Information Technology, Chinese Academy of Sciences, China\\
			$^{\S}$University of Chinese Academy of Sciences, Beijing 100049, China\\
			Email:  \{fumin, zhouyong, shiym\}@shanghaitech.edu.cn}
	}
	\maketitle
	
	\begin{abstract}
		In 	multiple-input multiple-output (MIMO) device-to-device (D2D) networks, interference and rank-deficient channels are the critical bottlenecks for achieving high degrees of freedom (DoFs). 
		In this paper, we propose a reconfigurable intelligent surface (RIS) assisted interference alignment strategy to simultaneous mitigate the co-channel interference and cope with rank-deficient channels, thereby improving the feasibility of interference alignment conditions and in turn increasing the achievable DoFs. 	
		The key enabler is a general low-rank optimization approach that  maximizes the achievable DoFs by jointly
		designing the phase-shift and transceiver matrices.  
		To address the  unique challenges of the coupled optimization variables, we develop a block-structured Riemannian pursuit method by solving  fixed-rank and unit modulus constrained least square subproblems along with rank increase. 
	    Finally, to reduce the computational complexity and achieve good DoF performance,
	    we develop unified  Riemannian conjugate gradient algorithms to alternately optimize  the fixed-rank transceiver matrix and  the unit modulus constrained phase shifter by exploiting the non-compact Stiefel manifold and the complex circle manifold, respectively.
		Numerical results  demonstrate the effectiveness of deploying an RIS and the superiority of the proposed block-structured Riemannian pursuit method in terms of the achievable DoFs and the achievable sum rate.

	\end{abstract}

\section{Introduction}\label{introduction}

Device-to-device (D2D) communication has been recognized as a promising technology  that can enhance the efficiency of wireless networks by allowing direct communication between devices in proximity  without going through base stations \cite{Asadi2014Survey}. 
However, interference is one of the critical issues in emerging D2D networks, where  the co-located transceiver pairs  may severely interfere with each other \cite{Shen2019Optimization}. 
Fortunately,
interference alignment  \cite{sridharan2015linear}  as a powerful  technology  has been well studied to manage interference in various interference-limited scenarios (e.g., multiple-input multiple-output (MIMO)  beamforming \cite{sridharan2015linear}).
The main idea  is to restrict the desired signal space  being a complement to the interference space. 
Such a linear interference management strategy achieves the optimal degrees of freedom (DoFs) of generic channels in high signal-to-noise ratio (SNR) \cite{sridharan2015linear}. 
However, due to poor scattering, insufficient antenna-spacing, and keyhole effects,  rank-deficient channels lead to  significant performance degradation in terms of achievable DoFs \cite{krishnamurthy2014degrees}.

Reconfigurable intelligent surfaces (RISs) as a cost-effective technology have recently been proposed to enhance the network performance \cite{ Yuan2020Reconfigurable, Rui_arXiv19IRSmag, yang2020federatedRIS}. 
Specifically, an RIS consists of a number of passive elements, each of which is able to dynamically re-scatter the incident signals  with the desired phase shift via a smart controller, thereby combating the unfavorable propagation conditions. 
There is a growing body of recent works  to study the performance of various RIS-assisted wireless networks.
The advantages of RIS were leveraged to reduce the energy consumption \cite{wu2018intelligent}, \cite{Fu2019Intelligent} and  increase the achievable data rate \cite{huang2018large, Guo2019Weighted, yu2019miso}.
In addition, various channel estimation methods were proposed to estimate channel coefficients of RIS-related links (e.g., compressive-sensing based methods \cite{Xia2019Intelligent}).
However, it is still not clear whether the feasibility of  interference alignment conditions can be  improved by RIS. 
This motivates us to  explore the benefits of  RIS  to improve the feasibility of interference alignment conditions  in D2D networks and in turn increase the DoFs without the assumption of independent channels.
  
In this paper, we consider an RIS-assisted MIMO D2D network, where the RIS is  proposed to assist interference alignment.
Specifically, we  jointly design the phase-shift matrix at RIS and the transceiver  matrix  to maximize the achievable DoFs. 
By establishing interference alignment conditions, we develop a rank minimization problem to maximize  DoFs. 
However, the formulated problem is highly intractable due to the non-convexity of the bilinear function, rank function, and the unit modulus constraints.
To address this unique challenges,  we propose a block-structured Riemannian pursuit method by  solving a sequence of non-convex  least square subproblems along with rank increase.
To deal with the coupled fixed-rank transceiver matrix  and unit modulus constrained phase shifters, we further develop unified 
low-complexity Riemannian conjugate gradient (RCG) methods with good performance \cite{Absil2009Optimization}, \cite{hu2019brief} to alternately optimize transceiver matrix and phase shifters  by exploiting the non-compact Stiefel manifold and complex circle manifold, respectively.
Simulation results validate the effectiveness of deploying an RIS and the performance gains of the proposed method in terms of the achievable DoFs and  the achievable sum rate.

$\textit{Organization}$: The remainder of this paper is organized as follows. Section II describes the system model and problem formulation. 
We present a block-structured Riemannian pursuit method  to solve the formulated problem in Section III. 
Section IV  presents the numerical results and Section V concludes this paper.
\section{System  Model and Problem Formulation}\label{model}
\subsection{System Model}
We consider an MIMO D2D network consisting of $K$  transceiver pairs with the assist of an RIS.
We denote $\mathcal{K} =\{1,\ldots,K  \}$ as the index set of D2D pairs.
The $k$-th transmitter and  receiver are equipped with  $N_k$ and $M_k$ antennas, respectively. 
With full frequency reuse, the D2D pairs may severely interfere with each other \cite{Asadi2014Survey}, \cite{Shen2019Optimization}. 
Hence, we propose to deploy an RIS equipped with  $L$ passive reflecting elements, aiming to  alleviate the severe co-channel interference among $K$ pairs. 
We denote $\bm \Theta = \text{diag}\{v_1, \ldots, v_L\} \in  \mathbb{C}^{L \times L}$ as the diagonal reflection matrix of the RIS. Specifically, let $v_l \in \mathbb{C}$ denote the reflection coefficient of the $l$-th RIS element, which is assumed to satisfy $|v_l| = 1, \forall l$ \cite{huang2018large, wu2018intelligent, Fu2019Intelligent,Guo2019Weighted, yu2019miso}.

We denote $\bm{H}_{ij} \in \mathbb{C}^{M_i \times N_j}$ as  the channel of direct link  between the  $i$-th receiver and the $j$-th transmitter,  $\bm R_i \in  \mathbb{C}^{ M_i\times L}$ as that between the RIS and the $i$-th receiver, and $\bm T_j \in  \mathbb{C}^{ L\times N_j}$
as that between  the $j$-th transmitter and the RIS.
Thus, the composite  channel matrix  between the  $i$-th receiver and the $j$-th transmitter, consisting of both the direct and reflect links, is given by $ \tilde{\bm H}_{ij} = \bm H_{ij} + \bm R_i\bm \Theta \bm T_j \in \mathbb{C}^{M_i \times N_j}$.

Each transmitter $i$ has a message $W_i$ intended for receiver $i$. Message $W_i$ is encoded into a vector $\bm{x}_{i} \in \mathbb{C}^{rN_i}$ of length $rN_i$ and transmitted by $N_i$ antennas over $r$ channel uses, where $\mathbb{E}[\bm{x}_{i}^{\sf H}\bm{x}_{i}] \leq P.$
 We consider the quasi-static fading  channels, i.e., $\bm H_{ij}$, $\bm R_i$, and $\bm T_j$  remain invariant over $r$ channel uses during the data transmission.  Hence, the phase-shift matrix at the RIS is designed to be  invariant during the data transmission.
 Hence, the signal received by the $m$-th antenna of receiver $i$ over $r$ channel uses  is given by
\setlength\arraycolsep{1.2pt}
\begin{eqnarray}
\bm{y}_i[m] &&= \sum_{j=1}^{K}\sum_{n=1}^{N_j}\Big(H_{ij}[m,n]+ \bm R_i[m]\bm \Theta \bm T_j[n] \Big)\bm{x}_{j}[n]+\bm{z}_i[m]\nonumber \\
                     &&= \sum_{j=1}^{K}\sum_{n=1}^{N_j}\tilde{H}_{ij}[m,n]\bm{x}_{j}[n]+\bm{z}_i[m],       
\end{eqnarray}
where $\bm{z}_i[m] \in \mathbb{C}^{r} \sim \mathcal{CN}(\bm{0},\sigma \bm{I}_{r})$ is the additive white Gaussian noise (AWGN), $\bm{x}_j[n]\in \mathbb{C}^{r}$ corresponds to the  signal transmitted by the $n$-th antenna at transmitter $j$  over $r$ channel uses, $\bm{y}_i[m]\in \mathbb{C}^{r}$ is the   signal received by the $m$-th antenna  at receiver $i$ over $r$ channel uses,  $\tilde{H}_{ij}[m,n]$ is the $(m,n)$-th entry of matrix $\tilde{\bm{H}}_{ij}$,  $H_{ij}[m,n]$ is the $(m,n)$-th entry of matrix $\bm{H}_{ij}$, $\bm R_i[m]  \in \mathbb{C}^{1\times L} $ is the $ m $-th row of matrix $ \bm R_i $, and $ \bm T_j[n] \in \mathbb{C}^{L}$ is the $ n $-th column of matrix $ \bm T_j $.

The  achieve rate $R(W_i)$ is achievable of message $W_i$  if there exists an encoding scheme such the message $W_i$ is $R(W_i)$ while the error probability of decoding $W_i$ for user $i$ can be made arbitrarily small  as the number of channel uses $r$ is  big enough.
For each message delivery, the DoF is characterized by the first order characterization of channel capacity and is defined by \cite{sridharan2015linear}, \cite{shi2016lowRP}
\begin{equation}
\textrm{DoF}(W_i)=\lim_{\text{SNR}_i\rightarrow\infty}\frac{R(W_i)}{\log(\text{SNR}_i)}, \forall i \in \mathcal{ K},
\end{equation}
where ${\text{SNR}_i}$ denotes the $i$-th transmitter's SNR.
We aim to design an effective linear interference alignment strategy to maximize the achievable DoFs in the sequel.

\subsection{Interference Alignment}
Linear interference alignment  have attracted considerable attention to manage the co-channel interference
due to their low-complexity and the DoF optimality in high SNR scenarios \cite{sridharan2015linear}. 
Specifically, let the linear precoding matrix at transmitter $i$  and decoding matrix at receiver $i$ over $r$ channel uses be $\bm{V}_i\in \mathbb{C}^{N_ir \times d_i}$ and $\bm{U}_i\in \mathbb{C}^{M_ir \times d_i}$, respectively. 
 Message $W_i$ is split into $d_i$ independent scalar data streams, denoted as $ \bm s_i \in \mathbb{C}^{d_i}$, and is transmitted along with the precoding matrix $\bm{V}_i$, i.e., $\bm x_i = \bm{V}_i \bm s_i$. 
We follow \cite{Shen2019Optimization}, and assume that the full CSI is  known and a centralized node  is responsible for the network optimization and then the feedback of the optimization parameters to the D2D pairs.
 Note that even with this ideal assumption, this kind of optimization problems is still NP-hard.
 Based on the aforementioned definitions, the received signal $\bm{y}_i\in\mathbb{C}^{rM_i}$ at receiver $i$ is given by
 \begin{eqnarray}\label{receive signal}
 {\bm{y}}_i=&&\left(\!{\tilde{\bm H}}_{ii}\otimes{\bm{I}}_r\right)\!{\bm{V}}_i{\bm{s}}_i \!+\!  \sum\nolimits_{i\ne j} \!\left({\tilde{\bm H}}_{ij}\otimes{\bm{I}}_r\!\right)\!{\bm{V}}_{j}{\bm{s}}_j \!+\! \bm{z}_{i},
 \end{eqnarray}
 where $\otimes$ is the Kronecker product operator and $\bm{z}_i \sim \mathcal{CN}(\bm{0},\sigma \bm{I}_{r M_i})$. 
 With decoding matrix $\bm U_i$, we have
 \begin{eqnarray}\label{decoding signal}
\!\! \tilde{\bm{y}}_{i}
 \!= \! \bm{U}_i^{\sf{H}}\!\left(\!{\tilde{\bm H}}_{ii} \!\otimes \!{\bm{I}}_r\right)\!{\bm{V}}_i{\bm{s}}_i \!+\! \bm{U}_i^{\sf{H}}  \sum \nolimits_{i\ne j} \!\left({\tilde{\bm H}}_{ij} \!\otimes \!{\bm{I}}_r\!\right)\!{\bm{V}}_{j}{\bm{s}}_j \!+\! \tilde{\bm{z}}_{i},
 \end{eqnarray}
 where $\tilde{\bm{y}}_{i} \in \mathbb{C}^{d_i}$, $\tilde{\bm{z}}_{i} = \bm U_i^{\sf H} \bm z_i\in \mathbb{C}^{d_i}$.
We observe that $\tilde{\bm{y}}_{i}$  can be decomposed into  desired signal, interference, and noise.
In the  high SNR regime, we exploit the interference alignment to design the precoding, decoding,  and phase-shifter matrices to  cancel interference while preserving the desired signals,
which imposes the following conditions \cite{sridharan2015linear} 
\begin{eqnarray}
\label{IA1}
{\bm{U}}_i^{\sf{H}}({\tilde{\bm H}}_{ij}\otimes {\bm{I}}_r){\bm{V}}_j&&={\bm{0}}, \forall i\ne j,\\
\label{IA2}
{\rm{rank}}\left({\bm{U}}_i^{\sf{H}}({\tilde{\bm H}}_{ii}\otimes {\bm{I}}_r){\bm{V}}_i\right)&&= d_i,
\forall i=1,\dots, K.
\end{eqnarray}
Equation \eqref{IA1} guarantees all the interfering signals at receiver $i$ lie in the subspace orthogonal to  $\bm U_i$, while Equation \eqref{IA2} assures that the signal subspace $({\tilde{\bm H}}_{ii}\otimes {\bm{I}}_r){\bm{V}}_i$ has dimension $d_i$ and is linearly independent of the interference subspace. 
If conditions \eqref{IA1} and \eqref{IA2} are satisfied, the parallel interference-free channels can be obtained over $r$ channel uses. 
Therefore, the DoF tuple $(d_1/r,\ldots,d_K/r)$ is achievable for messages $\{W_1,\ldots,W_K\}$. 
\subsection{Problem Formulation}
For fixed $\{d_1, \ldots, d_K\}$, the smaller the value of $r$,  the larger the achievable DoFs. 
As a result, we formulate a rank minimization problem to maximize the achievable DoFs in this subsection.
We first develop a low-rank model to find the minimal channel uses $r$ such that  the interference alignment conditions  \eqref{IA1} and \eqref{IA2} are feasible. 
We note that 
\begin{eqnarray}
\bm{U}_{i}^{\sf H}({\tilde{\bm H}}_{ij}\otimes\bm{I}_r)\bm{V}_{j}=\sum_{m=1}^{M_i}\sum_{n=1}^{N_j}\tilde{H}_{ij}[m,n]\bm{U}_{i}^{\sf H}[m]\bm{V}_{j}[n], 
\end{eqnarray}
 where $\bm{V}_{j}[n]\in\mathbb{C}^{r\times d_j}$ and $\bm{U}_{i}[m]\in\mathbb{C}^{r\times d_i}$ correspond to the $n$-th antenna of transmitter $j$ and the $m$-th antenna of receiver $i$  over $r$ channel uses, respectively.
We  define $\bm{X}_{i,j}[m,n]= \bm{U}_{i}^{\sf H}[m]\bm{V}_{j}[n] \in \mathbb{C}^{d_i \times d_j}$.
The rank constraints in \eqref{IA2} are represented by its column-reduced echelon form, i.e., $\bm{U}_{i}^{\sf H}(\tilde{\bm H}_{ii}\otimes\bm{I}_r)\bm{V}_{i}=\bm{I}$ \cite{sridharan2015linear}, \cite{shi2016lowRP} to support efficient algorithm design.
Thus, conditions  \eqref{IA1} and \eqref{IA2} can be rewritten as 
\begin{eqnarray}
\sum\nolimits_{m=1}^{M_i}\sum \nolimits_{n=1}^{N_j}\tilde{H}_{ij}[m,n]\bm{X}_{i,j}[m,n]&=&\bm{0},  \forall i \in \mathcal{K},  i\ne j,\label{IA11}\\
\sum\nolimits_{m=1}^{M_i}\sum\nolimits_{n=1}^{N_i}\tilde{H}_{ii}[m,n]\bm{X}_{i,i}[m,n]&=&\bm{I}, \forall i \in \mathcal{K}. \label{IA22}
\end{eqnarray}
Let  $M = \sum_{i=1}^{K}M_id_i$, $N = \sum_{j=1}^{K}N_jd_j$, and $S = \sum_{i=1}^{K}\sum_{j=1}^{K}d_id_j$. 
By defining a set of matrices
\begin{eqnarray*}
\tilde{\bm{V}}_{j}
&&=\begin{bmatrix}
\bm{V}_{j}[1], & \ldots &, \bm{V}_{j}[N_j] 
\end{bmatrix},
\tilde{\bm{U}}_{i}
=\begin{bmatrix}
\bm{U}_{i}[1], & \ldots&,  \bm{U}_{i}[M_i]
\end{bmatrix} ,\\
\tilde{\bm{V}}
&&=\begin{bmatrix}
\tilde{\bm{V}}_{1}, & \ldots &, \tilde{\bm{V}}_{K} 
\end{bmatrix},
\tilde{\bm{U}} =
\begin{bmatrix}
\tilde{\bm{U}}_{1}, & \ldots &, \tilde{\bm{U}}_{K} 
\end{bmatrix},
\end{eqnarray*}
 where $\tilde{\bm{V}}_{j}\in \mathbb{C}^{r\times N_jd_j}$, $\tilde{\bm{U}}_{i}\in\mathbb{C}^{r\times M_id_i}$, $\tilde{\bm{V}}_{1}\in \mathbb{C}^{r\times N}$, and $\tilde{\bm{U}}\in \mathbb{C}^{r\times M} $,    we further have 
 \vspace{-0.2cm}
\begin{eqnarray}
\bm{X}_{i,j} &&= [\bm{X}_{i,j}[m,n]] =  \tilde{\bm{U}}_{i}^{\sf H}\tilde{\bm{V}}_{j} \in \mathbb{C}^{M_id_i \times N_jd_j},\\
\bm{X}&&= [\bm{X}_{i,j}] = \tilde{\bm{U}}^{\sf H} \tilde{\bm{V}}\in \mathbb{C}^{M \times N}.
\end{eqnarray}
Note that the rank of matrix $\bm{X}$ is equal to the number of channel uses $r$ since $\bm{X}=\tilde{\bm{U}}^{\sf H}\tilde{\bm{V}}$, i.e., $\textrm{rank}(\bm{X})=r$. 
We  vectorize both sides of \eqref{IA11} and  \eqref{IA22},  followed by  characterizing both equations as  $\mathcal{A}(\bm{X}, \bm \Theta)=\bm{b}$ with the bilinear operator $\mathcal{A}:\mathbb{C}^{M\times N}\mapsto \mathbb{C}^{S}$ as a function of $\{\bm H_{ij}, \bm R_i, \bm T_j\}$.  
We hence propose the following generalized low-rank optimization problem to maximize the achievable  DoFs
\begin{eqnarray}
\mathscr{P}:\mathop{\textrm{minimize}}_{\bm{X},\bm \Theta} && \textrm{rank}(\bm{X}) \nonumber \\
\textrm{subject to} && \mathcal{A}(\bm{X}, \bm \Theta)=\bm{b},\nonumber\\
&& |v_l| = 1, \forall l = 1,\ldots, L.
\end{eqnarray}
 However, Problem $\mathscr{P}$ is computationally difficult due to the non-convexity of the rank function, the bilinear equation constraint,  and the unit modulus constraints. 
 General convex relaxations (e.g., nuclear norm for rank) for Problem $\mathscr{P}$ are inapplicable  due to the bilinear constraint and the unit modulus constraints. 
  In the following section, we propose a block-structured Riemannian pursuit method  to solve   $\mathscr{P}$ in the manifold space to achieve the algorithmic advantages and admirable performance.
  \begin{figure*}[t]
	\centering
	\subfigure[Tangent space and Riemannian gradient]{\includegraphics[width=0.65\columnwidth,height=3.5cm]{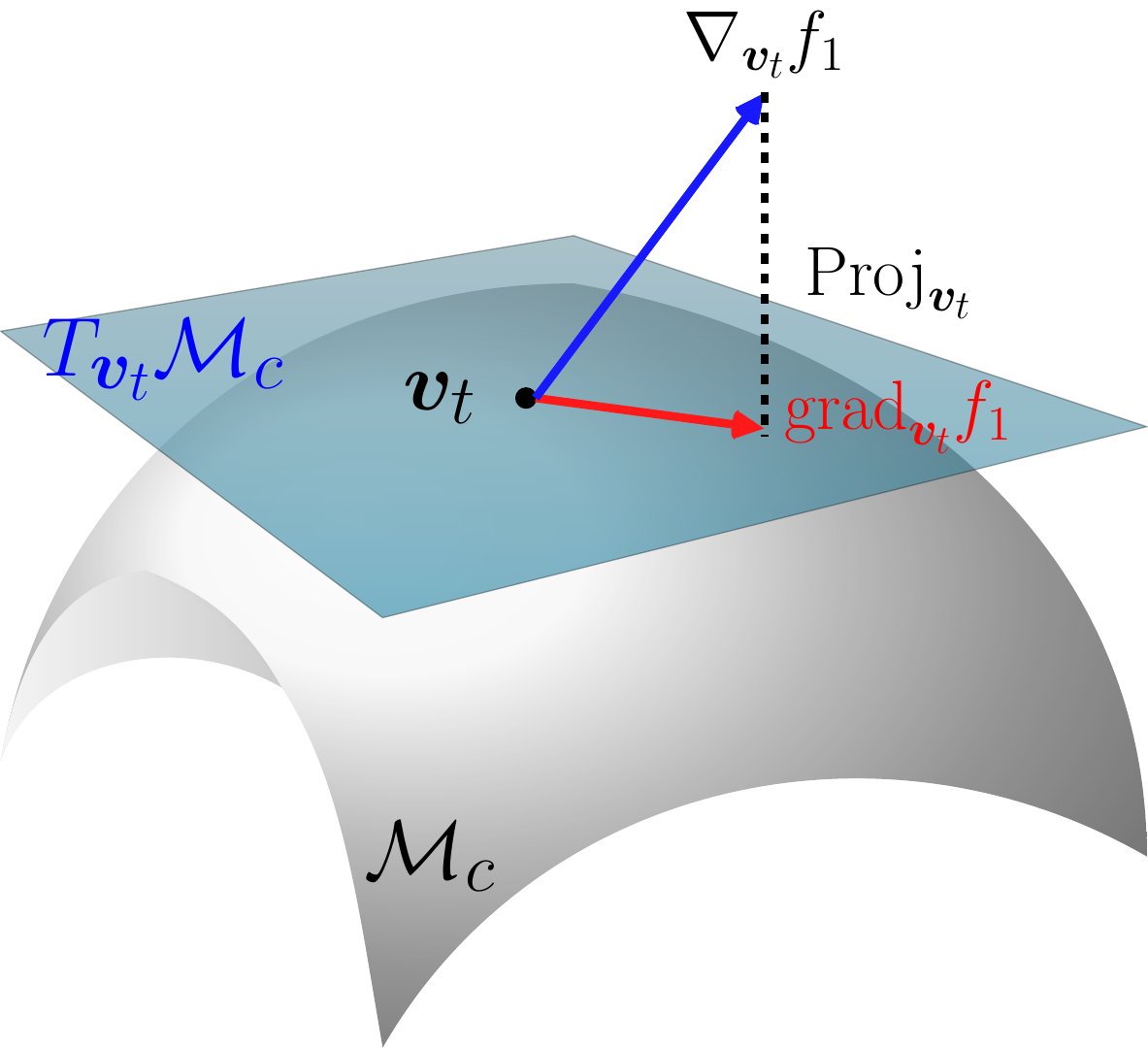}\label{fig:derivative1}}\hfil
	\subfigure[Vector Transport]{\includegraphics[width=0.65\columnwidth,height=3.5cm]{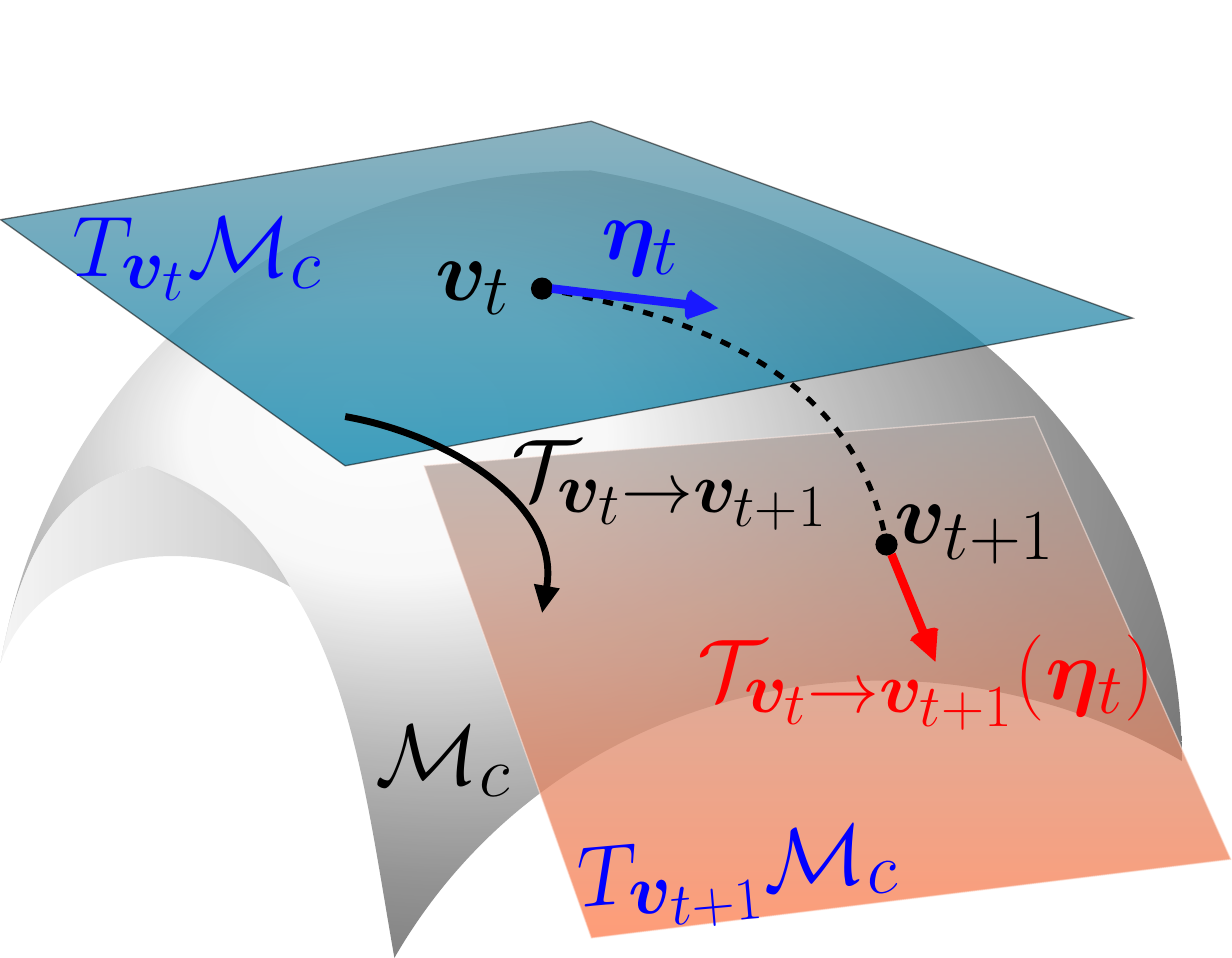}\label{fig:transport}}\hfil
	\subfigure[Retraction]{\includegraphics[width=0.65\columnwidth,height=3.5cm]{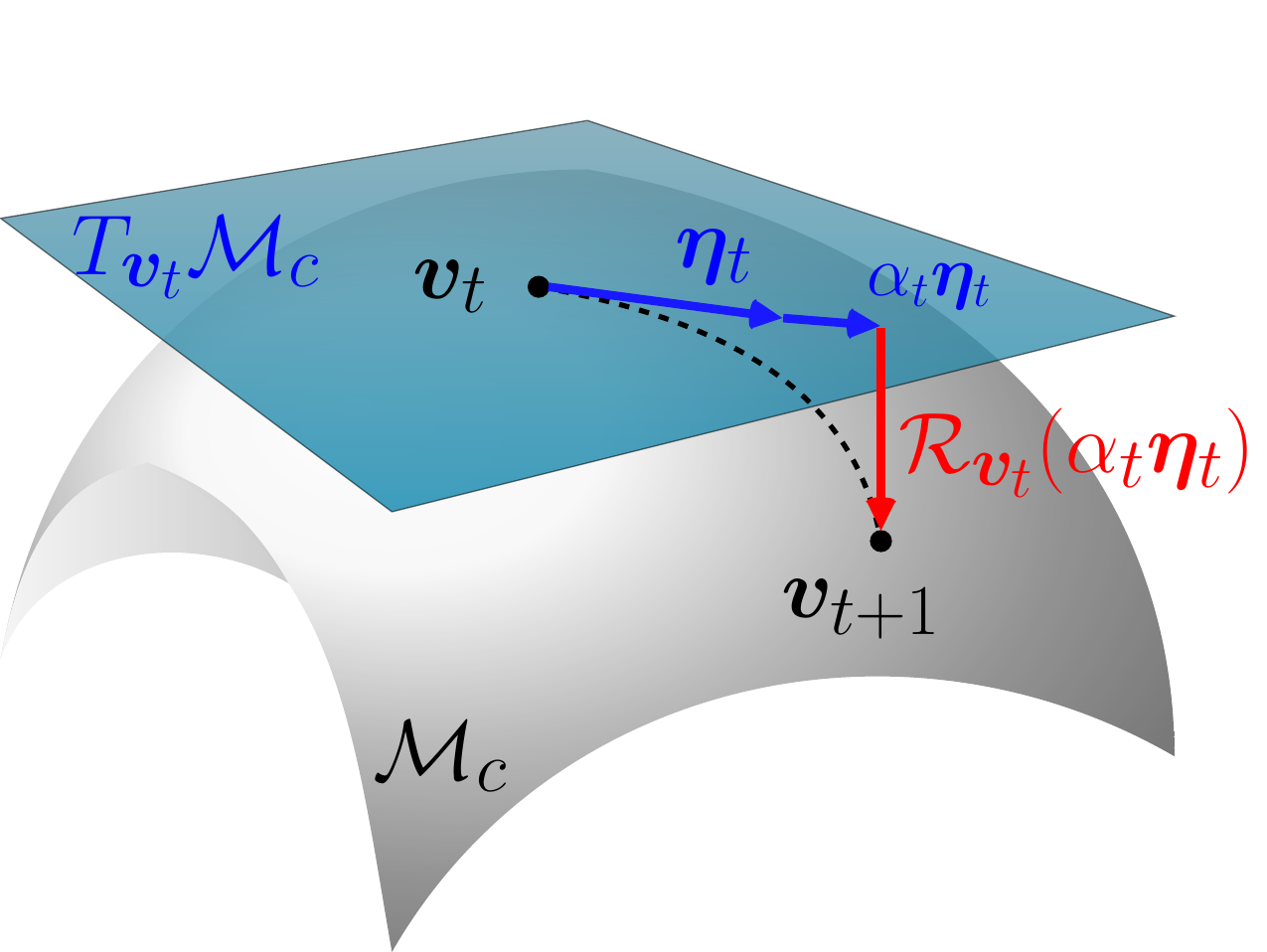}\label{fig:retraction}} 
	\caption{An illustrative example of the key steps in manifold optimization.}\label{manofold}
	\vspace{-0.4cm}
\end{figure*}
 \section{Block-Structured Riemannian Pursuit}\label{subsec:RP_algo}
 In this section, we develop a block-structured Riemannian pursuit for the rank  minimization  problem in the RIS-assisted MIMO D2D networks to reduce the computational complexity and achieve good performance.
\subsection{Rank Pursuit}
In this subsection, we present an efficient rank pursuit strategy to solve Problem $\mathscr P$ by alternately solving the fixed-rank optimization and rank increase, thereby detecting the minimum rank of matrix $\bm X$.
Specifically,  fixing the rank of matrix $\bm X$ as $r$, we need to solve the following non-convex  
optimization problem
\begin{eqnarray}\label{fix rank}
\mathop{\textrm{minimize}}_{\bm{X},\bm \Theta} && f_0(\bm X, \bm \Theta)=\frac{1}{2}\|\mathcal{A}({\bm X}, \bm \Theta)-{\bm{b}}\|_2^2 \nonumber\\
\textrm{subject to}&& \textrm{rank}({\bm X})=r,\nonumber\\
&& |v_l| = 1, \forall l = 1,\ldots, L.
\end{eqnarray}
Although the problem $\eqref{fix rank}$ is still non-convex, we observe that the objective function $f_0$ is smooth about each  block variable which stays different manifold spaces, i.e., fixed-rank complex matrix in non-compact Stiefel manifold and unit modulus constrained phase shifters in the complex circle manifold.
The low computational complexity and good performance of Riemannian optimization \cite{Absil2009Optimization}, \cite{ hu2019brief}  have  been validated in various wireless networks  e.g., topological interference management  \cite{shi2016lowRP,Shi2017admission,Yang2019Generalized}, and blind demixing \cite{Dong2019Blind}. 
Hence, we  decouple the variables into two blocks, following by  adopting unified RCG algorithms to design the transceiver and phase-shift matrices on the manifold space alternately.  
The proposed algorithm is summarized in \textbf{Algorithm \ref{BSRPalgo}}.
\begin{algorithm}[t]
	\caption{Block-Structured Riemannian Pursuit for Solving Problem $\mathscr{P}$}
	\begin{algorithmic}[1]
		\STATE {\textbf{Input}}: $M$,  $N$, $L$, and desired
		accuracy $\epsilon$.
		\WHILE{not converged}
		\STATE Initialize: ${\bm{X}}_{0}^{[r]}\in\mathbb{C}^{M\times N}$, $\bm {\Theta}_0^{[r]}\in\mathbb{C}^{L\times L}$.
		\REPEAT
		\STATE Compute a critical point ${\bm{X}}_{t}^{[r]}$ for the smooth rank-$r$
		problem $\mathscr{P}_{2}$ with initial point ${\bf{X}}_{t-1}^{[r]}$ and $\bm {\Theta}_{t-1}^{[r]}$ using the Riemannian 
		algorithm in \cite{Yang2019Generalized}. 	
		\STATE Compute a critical point ${\bm{\Theta}}_t^{[r]}$ for the smooth unit modulus constrained
		problem $\mathscr{P}_{1}$ 
		with initial point $\bm {\Theta}_{t-1}^{[r]}$ and ${\bm{X}}_{t}^{[r]}$  using Algorithm \ref{XRCGalgoTheta}. 	
		\STATE Update $t=t+1$. 
		\UNTIL $\mathscr{P}_{0}\leq \epsilon$  or the maximum number of iterations $T$ is reached.
		\STATE Update the rank $r\leftarrow r+1$.  
		\ENDWHILE 
		\STATE {\textbf{Output}}: ${\bf{X}}^{[r]}$,  $\bm {\Theta}^{[r]}$  and the detected minimum rank $r$.
	\end{algorithmic}
	\label{BSRPalgo} 
\end{algorithm}     
\subsection{Optimizing Block $\bm \Theta$}
We define $\bm v = [v_1^{\sf H}, \ldots, v_L^{\sf H}]^{\sf H}\in \mathbb{C}^{L}$.  By denoting $\bm B_{ij}[m,n] = H_{ij}[m,n]\bm{x}_{ij}[m,n]\in \mathbb{C}^{d_i \times d_j}$ and $ \bm a_{ij}[m,n] = \bm R_i[m]\text{diag}(\bm T_j[n])  \in  \mathbb{C}^{1\times L}$, given fixed block $\bm X$, we have 
\begin{eqnarray}
\Big(H_{ij}[m,n]+ \bm R_i[m]\bm \Theta \bm T_j[n] \Big)\bm{X}_{ij}[m,n] \nonumber
\\
= \bm B_{ij}[m,n] +\bm a_{ij}[m,n] \bm v \bm{X}_{ij}[m,n]. 
\end{eqnarray}
The  conditions \eqref{IA11} and \eqref{IA22} can be rewritten as 
\begin{eqnarray}
\!\!\!\!\!\!\sum_{m=1}^{M_i}\sum_{n=1}^{N_j}\bm B_{ij}[m,n] \!+\!\bm a_{ij}[m,n] \bm v \bm{X}_{ij}[m,n]&&\!=\!\bm{0}, \forall i, i\ne j,\label{thetaIA11}\\
\!\!\!\!\!\!\sum_{m=1}^{M_i}\sum_{n=1}^{N_i}\bm B_{ii}[m,n] \!+\!\bm a_{ii}[m,n] \bm v \bm{X}_{ii}[m,n]&&\!=\!\bm{I}, \forall i. \label{thetaIA22}
\end{eqnarray}
By vectorizing  \eqref{thetaIA11} and \eqref{thetaIA22}, we have
\begin{eqnarray}
&&\text{Vec}\Big(\sum\nolimits_{m=1}^{M_i}\sum\nolimits_{n=1}^{N_j}\bm B_{ij}[m,n] +\bm a_{ij}[m,n] \bm v \bm{X}_{ij}[m,n]\Big) \nonumber \\
 && = \Big(\sum\nolimits_{m=1}^{M_i}\sum\nolimits_{n=1}^{N_j}\big( \text{Vec}(  \bm{X}_{ij}[m,n])\otimes \bm a_{ij}[m,n]\big) \Big)\bm v  + \nonumber\\&& \ \ \ \ \ \ \ \sum\nolimits_{m=1}^{M_i}\sum\nolimits_{n=1}^{N_j}\text{Vec}\big(\bm B_{ij}[m,n]\big). 
\end{eqnarray}
Hence, problem \eqref{fix rank} can be formulated as follows
\begin{eqnarray}\label{optiV}
\mathscr{P}_1: \mathop{\textrm{minimize}}_{\bm{v}} && f_1(\bm v)=\frac{1}{2}\|\bm{A}\bm v-{\bm{c}}\|_2^2 \nonumber\\
\textrm{subject to} 
&& |v_l| = 1, \forall l = 1,\ldots, L,
\end{eqnarray}
where
\begin{align*}
\bm {A}_{ij}& \!=\!\sum\nolimits_{m=1}^{M_i}\!\sum\nolimits_{n=1}^{N_j}\!\Big(\! \text{Vec}(  \bm{X}_{ij}[m,n])\!\otimes\! \bm a_{ij}[m,n]\!\Big) \!\in\!\mathbb{C}^{d_id_j\!\times\! L} ,\\
\!\bm A &\!=\! [\bm {A}_{11};\bm {A}_{12};\ldots;\bm {A}_{ij};\ldots; \bm {A}_{KK} ]\in  \mathbb{C}^{S\times L},\\
\!\bm {e}_{ij} &\!=\! \sum\nolimits_{m=1}^{M_i}\sum\nolimits_{n=1}^{N_j}\text{Vec}\big(\bm B_{ij}[m,n]\big)\in  \mathbb{C}^{d_id_j \times 1},\\
\!\bm {e} &\!=\! [\bm {e}_{11};\ldots; \bm {e}_{ij};\ldots;\bm {e}_{KK} ]\in  \mathbb{C}^{S\times 1},  \bm c= \bm b - \bm e\in  \mathbb{C}^{S\times 1}.
\end{align*}

Problem $\mathscr{P}_1$  is a non-convex quadratically constrained quadratic programming.  
The non-convex unit modulus constraints are the main obstacles of solving $\mathscr{P}_1$.  
Note that $\bm v$ lies on the manifold encoded by product of $L$ circles in the complex plane, denoted as $\mathcal{M}_c = \{ \bm v \in \mathbb{C}^{L}: |v_1| =  \cdots = |v_L| = 1   \}$ \cite{manopt2014}. 
We aim to develop computationally efficient Riemanian optimization algorithm on the product manifold of $L$ complex circles.

Specifically, the main idea  is to generalize a conjugate gradient (CG) method from Euclidean space to manifold space.
Similarly, we need to compute search directions in the tangent space and appropriate stepsizes on the manifold.
The main steps of the RCG algorithm consists of three steps in each iteration as shown in Fig. \ref{manofold}.
\subsubsection{Tangent Space and Riemannian Gradient}
A tangent space  $T_{\bm{v}_t}\mathcal{M}_c$ is composed of all the tangent vectors to $\mathcal{M}_c$ at any point $\bm{v}_t$ on a manifold. The Riemannian gradient is one tangent vector (direction) with the decrease of the objective function over the manifold space. 
For the complex circle manifold $\mathcal{M}_c$, the tangent space at $\bm{v}_t$ is given by
\begin{eqnarray}
T_{\bm{v}_t}\mathcal{M}_c =\left\{\bm{z}\in\mathbb{C}^L:\Re\left\{\bm{z}\odot\bm{v}_t^{*}\right\}=\bm{0}_L\right\}.
\end{eqnarray}
As shown in Fig. \ref{fig:derivative1}, the Riemannian gradient of  $f_1$ at $\bm{v}_t$, denoted by $\mathrm{grad}_{\bm{v}_t}f_1$, can be obtained by  orthogonally projecting the Euclidean gradient $\nabla_{\bm{v}_t}f_1$ onto the tangent space  $T_{\bm{v}_t}\mathcal{M}_c$ given by
\begin{eqnarray}\label{rvgradient}
\mathrm{grad}_{\bm{v}_t}f_1=\nabla_{\bm{v}_t} f_1-\Re\{\nabla_{\bm{v}_t} f_1\odot \bm{v}_t^{*}\}\circ\bm{v}_t,
\end{eqnarray}
where the Euclidean gradient of $f_1$ at $\bm{v}_t$ is given by
\begin{eqnarray}\label{evgradient}
\nabla_{\bm{v}_t} f_1 = \bm{A}^{\sf H}\left(\bm{A}\bm v_t-\bm{c}\right).
\end{eqnarray}
\subsubsection{Transport}
The search directions $\boldsymbol{\eta}_t$ and $\boldsymbol{\eta}_{t+1}$ in manifold optimization generally lie in two different tangent spaces $T_{\bm{v}_t}\mathcal{M}_c$ and $T_{\bm{v}_{t+1}}\mathcal{M}_c$, respectively. 
Therefore, the vector transport $\mathcal{T}_{\bm{v}_t\to\bm{v}_{t+1}}\left(\boldsymbol{\eta}_t\right)$
for manifold $\mathcal{M}_c$ is  the map of a tangent vector $\boldsymbol{\eta}_t$  from $T_{\bm{v}_t}\mathcal{M}_c$ to   $T_{\bm{v}_{t+1}}\mathcal{M}_c$ given by, as shown in Fig. \ref{fig:transport}, 
\begin{eqnarray}\label{transport}
\mathcal{T}_{\bm{v}_t\to\bm{v}_{t+1}}\left(\boldsymbol{\eta}_t\right)\triangleq \boldsymbol{\eta}_t\mapsto \boldsymbol{\eta}_t-\Re\{\boldsymbol{\eta}_t\odot \bm{v}_{t+1}^*\}\odot\bm{v}_{t+1}.
\end{eqnarray}
Thus, the update rule of the search direction for the RCG  is 
\begin{equation}\label{eq17}
\boldsymbol{\eta}_{t+1}=-\mathrm{grad}_{\bm{v}_{t+1}}f_1+\beta_t\mathcal{T}_{\bm{v}_t\to\bm{v}_{t+1}}\left(\boldsymbol{\eta}_t\right),
\end{equation}
where $\beta_t$ is chosen as the Polak-Ribiere parameter \cite{Absil2009Optimization}.
\subsubsection{Retraction}
After determining the search direction $\boldsymbol{\eta}_t$ at $\bm{v}_t$ and armijo backtracking line search step size $\alpha_t$ that the obtained point $\alpha_t\boldsymbol{\eta}_t$ does not lie in $\mathcal{M}_c$, we need to map it from the tangent space $T_{\bm{v}_t}\mathcal{M}_c$ to the manifold $\mathcal{M}_c$ by using retraction operator given by, as shown in Fig. \ref{fig:retraction},
\begin{equation}\label{retraction}
\bm{v}_{t+1} = \mathcal{R}_{\bm{v}_t}\left(\alpha_t\boldsymbol{\eta}_t\right)\triangleq
\alpha_t\boldsymbol{\eta}_t\mapsto \left(\alpha_t\boldsymbol{\eta}_t\right) \odot \frac{1}{\left(\alpha_t\boldsymbol{\eta}_t\right)} .
\end{equation}

Now, the key steps used in each iteration of the manifold optimization have been introduced. The resulting algorithm for solving $\mathscr{P}_1$ is summarized in \textbf{Algorithm \ref{XRCGalgoTheta}}.

 \subsection{Optimizing Block $\bm X$ }
 For a given phase-shifter matrix $ \bm\Theta $,  the concatenated channel response $\tilde{\bm H}_{ij}= \bm H_{ij} + \bm R_i \bm \Theta \bm T_j$ is fixed, and hence problem \eqref{fix rank} can be simplified as
   \begin{eqnarray}\label{fixRIS}
  \mathop{\textrm{minimize}}_{\bm{X}} && f_2(\bm X)=\frac{1}{2}\|\mathcal{A}_2({\bm X})-{\bm{b}}\|_2^2 \nonumber\\
   \textrm{subject to}&& \textrm{rank}({\bm X})=r,
   \end{eqnarray}           
 where $\mathcal{A}_2$ is a linear operator: $\mathbb{C}^{M\times N}\mapsto \mathbb{C}^{S}$ as a function of $\{\tilde{\bm H}_{ij}\}$.  
 This is a classical low-rank matrix completion (LRMC) problem, which has been studied extensively in the literature using convex and non-convex approaches.  The authors in \cite{Yang2019Generalized} proposed a Riemannian optimization algorithm on manifold with  better performance and higher computational efficiency compared with that of in Euclidean space  for topological cooperation, where the problem \eqref{fixRIS} is reformulated into the following problem
\begin{eqnarray}
 \mathscr{P}_2: \mathop{\textrm{minimize}}_{\bm{Y}\in \mathbb{C}^{N\times r}} && f_2(\bm Y)=\frac{1}{2}\|\mathcal{B}_2(\bm Y \bm Y^{\sf{H}})-\bm{b}\|_2^2 .
\end{eqnarray}  
where ${\bm{X}}={\bm{L}}{\bm{R}}^{\sf{H}}$, $\bm Y= \begin{bmatrix}
\bm L ; \bm R \end{bmatrix}\in \mathbb{C}^{(M+N)\times r}$\, and $\mathcal{B}_2(\bm Y \bm Y^{\sf{H}} ) = \mathcal{A}_2({\bm L \bm R^\mathsf{H}})$.
This is a Riemannian optimization problem with a smooth objective function over the complex non-compact Stiefel manifold $\mathcal{M}_r = \{\bm Y\in \mathbb{C}^{(M+N)\times r}: \text{rank}(\bm Y) = r\}$, i.e., the set of all $N\times r$ full column rank matrices in complex field. Similar to \textbf{Algorithm \ref{XRCGalgoTheta}}, the details of the RCG algorithm for the Problem $\mathscr P_2$ are shown in \cite{Yang2019Generalized}. We use the manifold optimization toolbox Manopt \cite{manopt2014} to implement the proposed RCG algorithms.
 \begin{algorithm}
 	\caption{RCG Algorithm for  Problem
 		$\mathscr{P}_{1}$}
 	\begin{algorithmic}[1]
 		\STATE {\textbf{Input}}: $L$, desired
 		accuracy $\varepsilon$.
 		\STATE Initialize: ${\bm{v}}_0= {\bm v}^{\textrm{initial}}, {\bm{\eta}}_0=-\mathrm{grad}_{\bm{v}_0 }f_1,
 		t=0$.
 		\REPEAT
 		\STATE Compute the search direction ${\bm{\eta}}_{t}\in T_{{\bm{	v}}_{t}}\mathcal{M}_{c}$ \\ according to \eqref{eq17}.
 		\STATE Update ${\bm{v}}_{t+1}={\mathcal{R}}_{{{\bm{v}}_{t}}}(\alpha_t{\bm{\eta}}_t)$ according to \eqref{retraction}.	
 		\STATE Update $t=t+1$. 
 		\UNTIL $\|\mathrm{grad}_{\bm{v}_t}f_1\|_2 \leq \varepsilon$ or the maximum number of iterations is reached.
 		\STATE {\textbf{Output}}: ${\bm{v}}^{\star}={\bm{v}}_{t}$, ${\bm{\Theta}}^{\star}=\text{diag}({\bm{v}}^{\star})$.
 	\end{algorithmic}
 	\label{XRCGalgoTheta}
 \end{algorithm}  
\section{Numerical Results}
\begin{figure*}[t]
	\centering
	\subfigure[Achievable DoF vs. The number of receive antennas $M_1$.]{\includegraphics[width=1.7in,height=4cm]{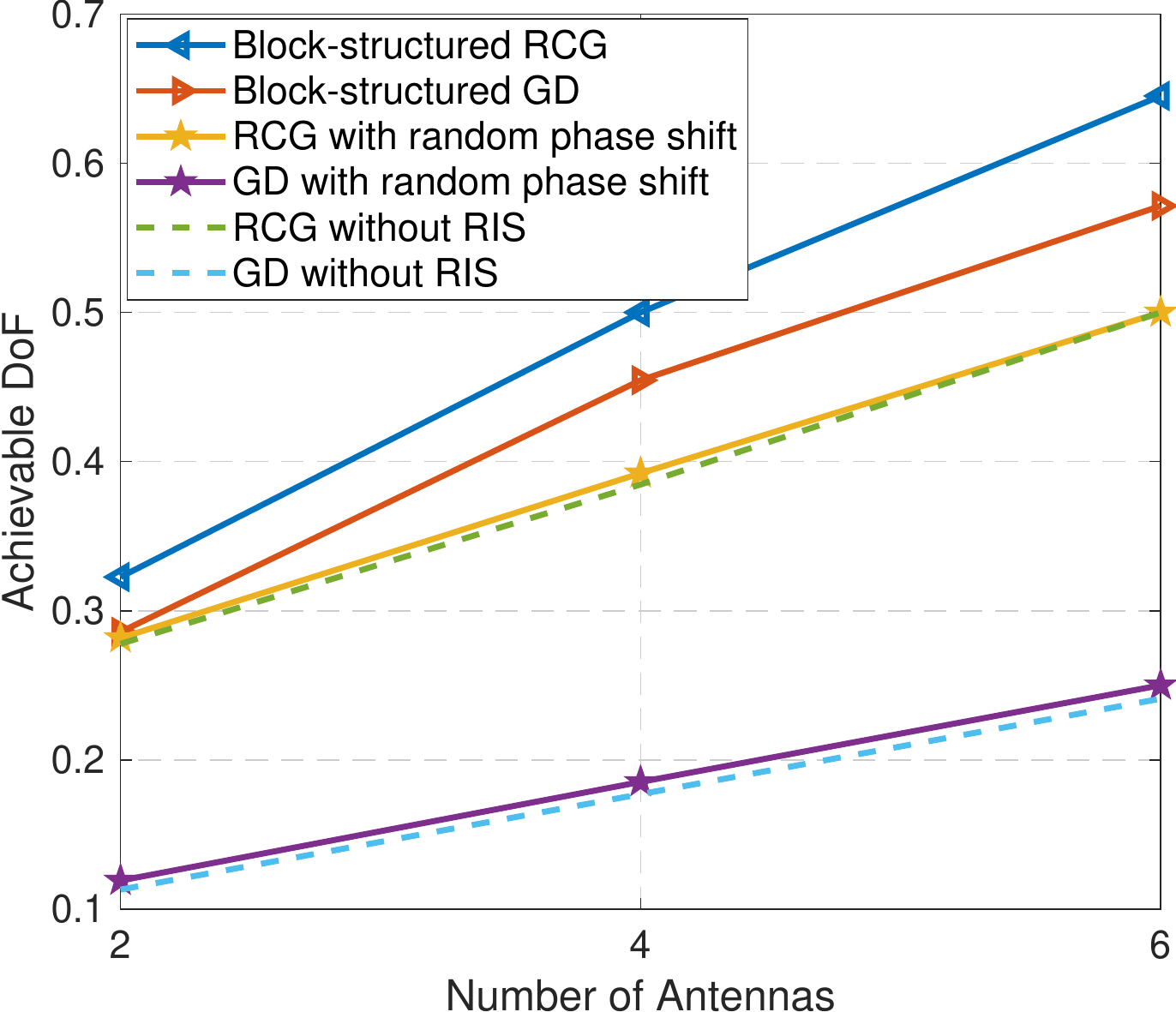}\label{fig1}}	
	\subfigure[Achievable sum rate vs. SNR at the transmitter.]{\includegraphics[width=1.7in,height=4cm]{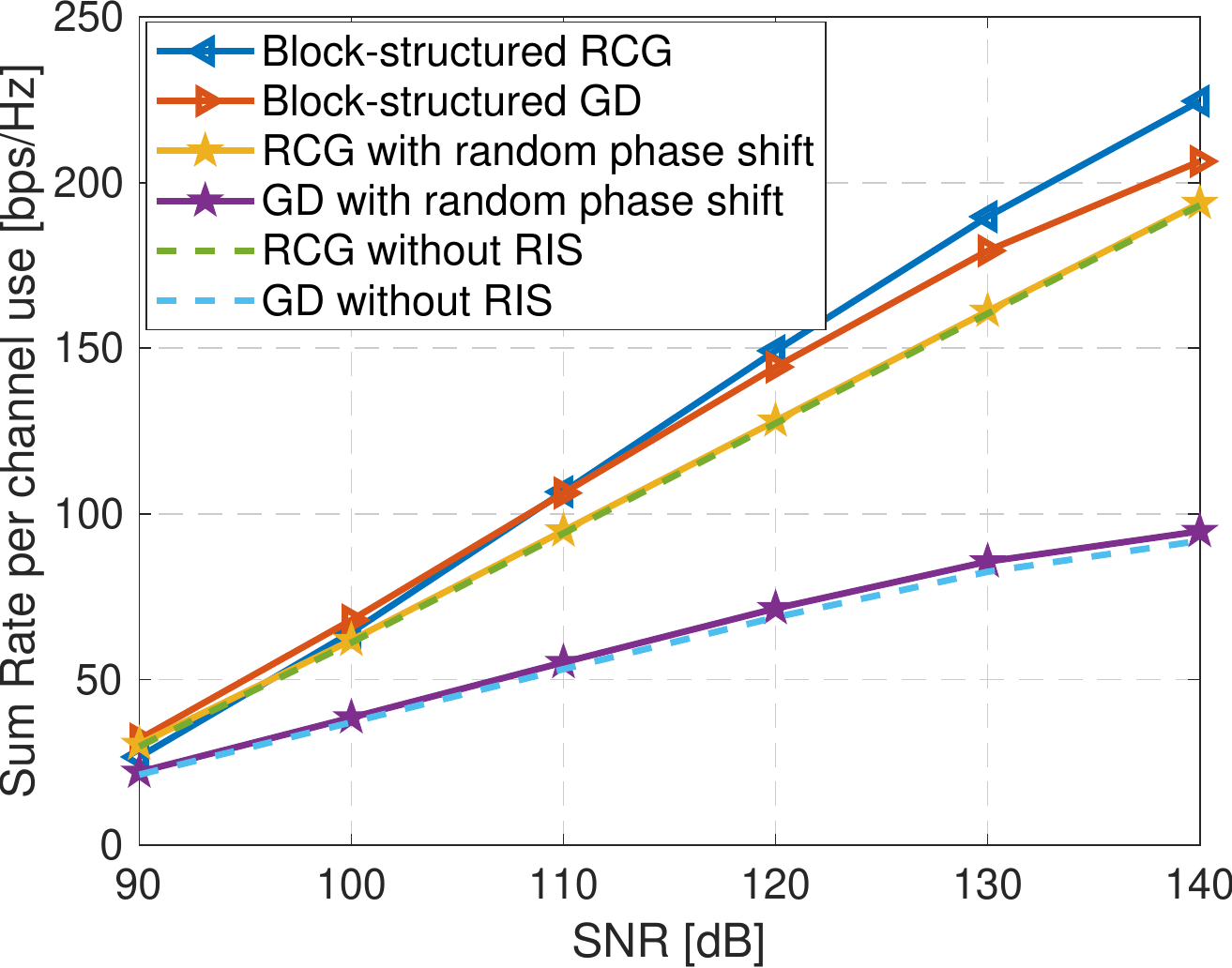}\label{fig2}}
	\subfigure[Achievable DoF  vs. rician factor $\beta_{RT}$ .]{\includegraphics[width=1.7in,height=4cm]{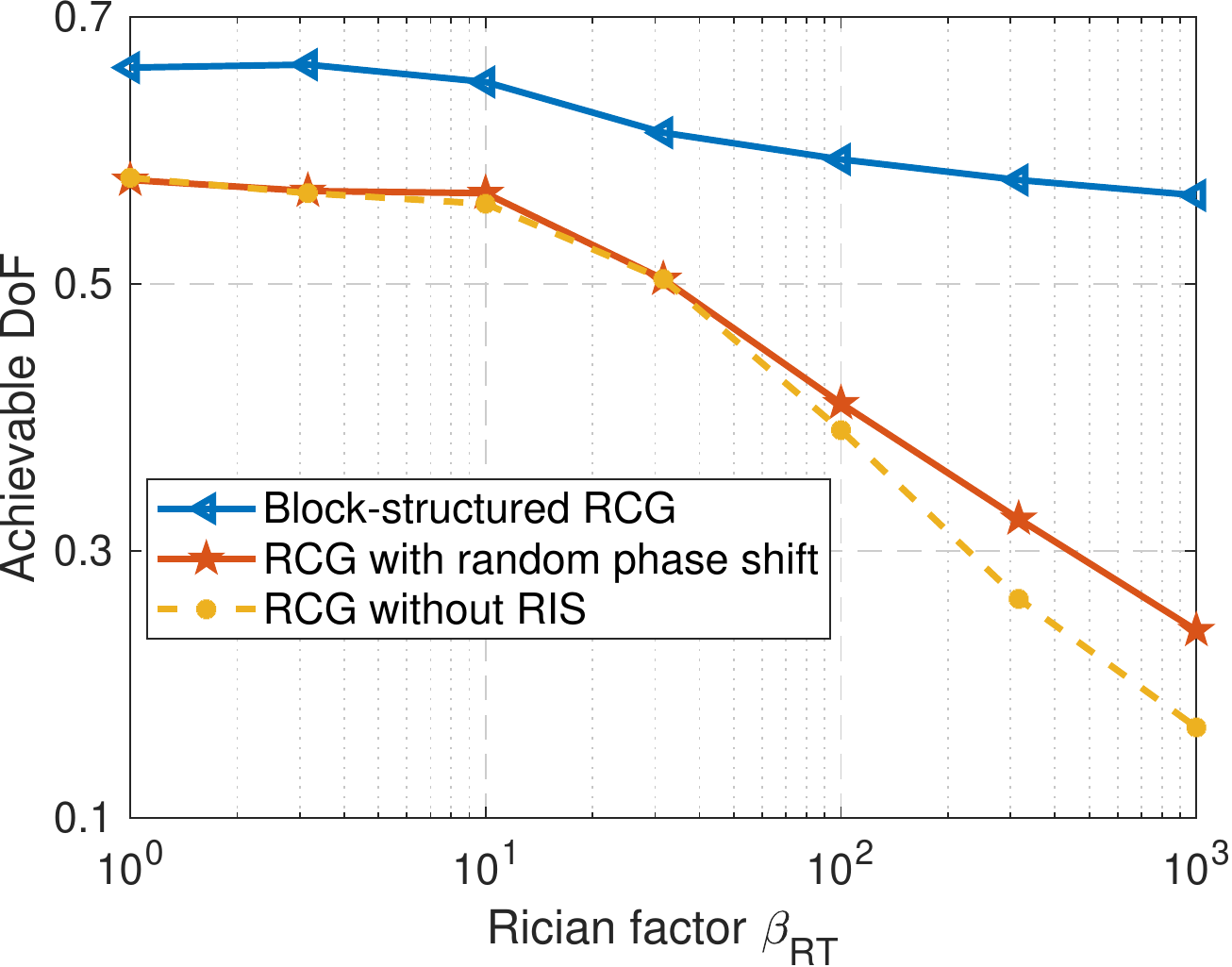}\label{fig3}}
		\subfigure[Achievable sum rate  vs. The number of RIS elements $L$ .]{\includegraphics[width=1.7in,height=4cm]{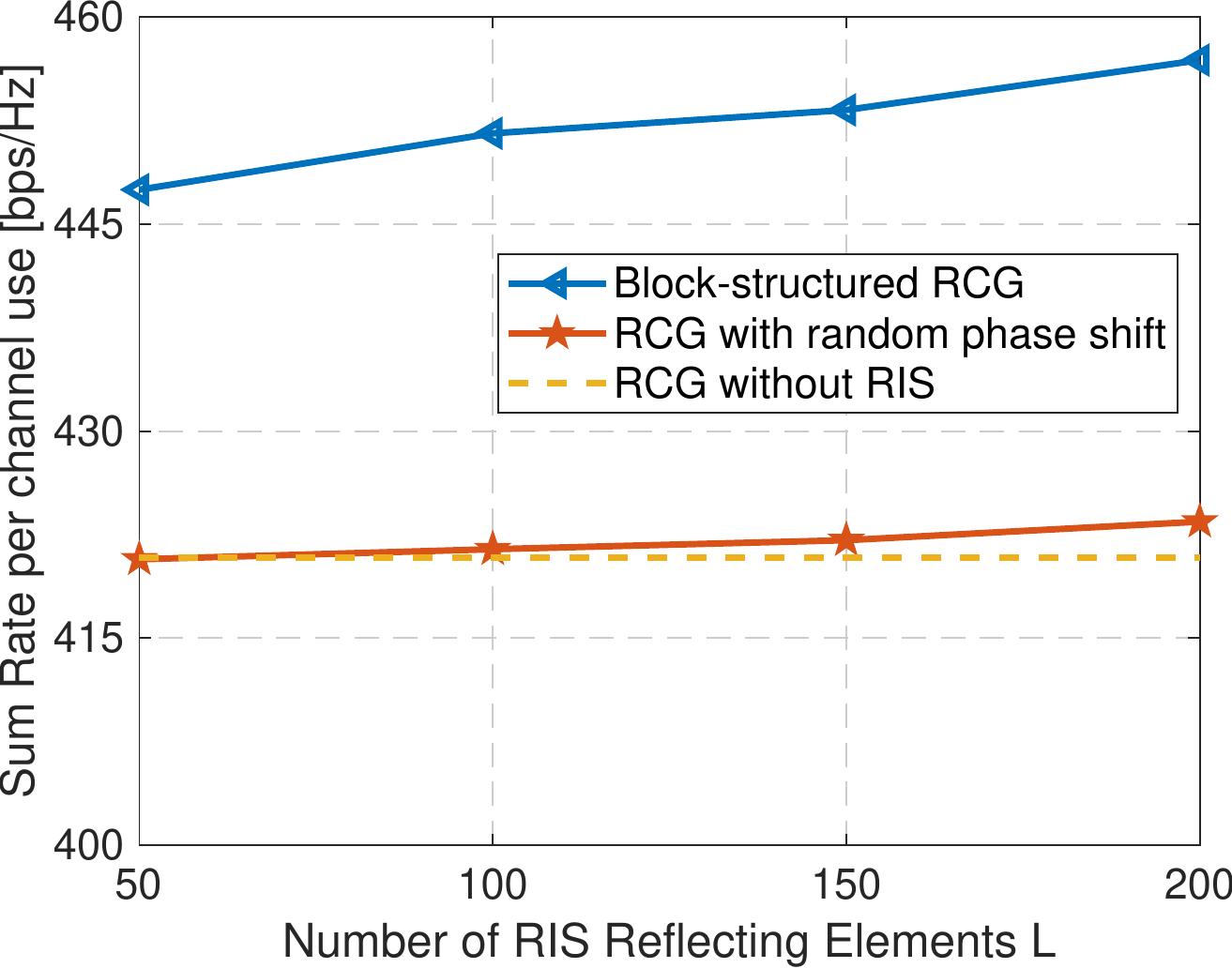}\label{fig4}}
	\caption{Performance comparisons between the proposed method and state-of-the-art methods under different network settings.}\label{fig}
    \vspace{-0.4cm}
\end{figure*}

In this section, we present the numerical results of the proposed   block-structured Riemannian pursuit method  in RIS-assisted MIMO D2D networks.
 We assume that  RIS is placed at $(25, 20)$ meters. 
The transmitters and receivers  are  randomly distributed in the region of $(0,20)\times(0,20)$ meters and $(30,50)\times(0,20)$ meters, respectively, the path loss model  is
$L(d) = T_0\left(d\right)^{-\alpha}$,
where $T_0=-30$ dB is the path loss at  reference distance, $d$ is the link distance, and $\alpha$ is the path loss exponent. 
The path loss exponents for the user-user link, the transmitter-RIS link, and the RIS-receiver link are set to 2.8, 2, and 2, respectively \cite{wu2018intelligent}.  
We denote $d_{\mathrm{RT}}^{ij}$, and $d_{\mathrm{IR}}^{i}$, $d_{\mathrm{IT}}^{j}$ as the distances between the $i$-th receiver and the $j$-th transmitter, between the $i$-th receiver and the RIS, and  between the $j$-th transmitter and the RIS, respectively. 
All channels are assumed to suffer from Rician fading \cite{ wu2018intelligent}. 
Hence, the corresponding channel coefficients are 
$$
\!\!\!\!\!\!\!\bm H_{ij} \!=\! \sqrt{L(d_{\mathrm{RT}}^{ij})}\!\left(\!\sqrt{\frac{\beta_{RT}}{1\!+\!\beta_{RT}}}  \bm H_{ij}^{\text{LOS}} \!+\!\sqrt{\frac{1}{1\!+\!\beta_{RT}}}  \bm H_{ij}^{\text{NLOS}}\!\right)\!,
$$
where $\beta_{RT}$ is the Rician factor, and $\bm H_{ij}^{\text{LOS}}$ and $\bm H_{ij}^{\text{NLOS}}$ denote the deterministic LoS and Rayleigh fading components, respectively.  The Rician factors of   the RIS-receiver link and the transmitter-RIS link  are denoted by $\beta_{IR}$ and $\beta_{IT}$, respectively.  All transmitters and receivers are equipped with $M_1$ and $N_1$ antennas, respectively. We set the number of data streams as $d_i=N_1,\forall i\in \mathcal{K}$.
We set $\sigma^2 = -120$ dB, $\epsilon = 10^{-4}$, $\varepsilon = 10^{-10}$, $T=30$, $K=10$, and $\beta_{IR}=\beta_{IT}=\beta_{RT}=10$.

For performance comparison, we consider the following schemes in simulations: 1) Block-structured RCG: Joint transceiver  and  phase-shift matrices design by
RCG proposed in Section III. 2) Block-structured GD: Joint  transceiver  and  phase-shift matrices by gradient descent algorithm (GD) given in \cite{Yang2019Generalized}. 3) Benchmark scheme with random phase shifts: Randomly choose the phase shifts.
4) Benchmark scheme without RIS.

Fig. \ref{fig1} plots the achievable DoF versus the number of receive antennas $M_1$ when $N_1 =2$ and $L = 150$. 
First, we observe that deploying more antennas at the receiver gives rise to the achievable DoF for all algorithms, due to the higher diversity and power gains. 
Furthermore, joint  transceiver  and  phase-shift matrices significantly  outperforms random phase shifts schemes. 
It demonstrates the necessity of  optimizing the phase-shifter matrix at the RIS, which enhance the conditionedness  of the composite channel matrices, thereby improving the feasibility of interference alignment conditions.
In addition, due to the superiority of the proposed RCG algorithms, the block-structured RCG algorithm can achieve higher DoFs than the block-structured GD algorithm.

In Fig. \ref{fig2}, we  investigate the achievable sum rate performance over different SNR values  when $N_1=2, M_1 = 6$, and $L=150$.   We observe that the sum rate can also be increased by jointly designing transceiver  and  phase-shift matrices, due to the DoF gain.
In addition, we observe that the block-structured RCG algorithm outperforms the block-structured GD algorithm in terms of the achievable sum rate. 

In Fig. \ref{fig3}, we plot the achievable DoFs versus the Rician factor of the transmitter-receiver channels $\beta_{RT}$ when $N_1 = M_1= 4$, and $L=50$ by using proposed algorithms shown in Section III.
One can observe that when $\beta_{RT}$ increases, the achievable DoFs  of the schemes without  RIS fall sharply. This is because for schemes without RIS, a higher Rician factor of $\bm H_{ij}$ results
in rank-deficient MIMO  channels, which leads to spatial multiplexing gain decrease.
However,  we observe that  deploying an RIS  make the achievable DoFs decrease slowly as  $\beta_{RT}$ increase.
This result is due to the fact that the reflected path by deploying the RIS makes angular separation at the receiver and thus provides  DoF gain. 
 Furthermore,  the achievable DoFs can be further enhanced to jointly design transceiver  and  phase-shift matrices  by using the block-structured RCG algorithm even when the number of passive reflecting elements is small.
The implication of this result is that it is  favorable to  optimize the phase shifters to compensate the direct-link rank-deficient MIMO  channels, so as to serve more multiple pairs that require sufficient DoFs compared to conventional D2D networks.

Fig. \ref{fig4} illustrates  the effect of the number of passive reflecting elements at the RIS (i.e., $L$) on the sum rate  when $N_1 =M_1 = 4$ and $\mathrm{SNR}=130$ dB.  
The achievable sum rate increases  as the value of $L$ increases by jointly designing the  transceiver matrix and phase-shift matrix.
 This result suggests the sum rate should be further improved by appropriately setting  the number of RIS elements while decreasing the transmitted power.

\section{Conclusion}\label{conclusion}
In this paper, we considered an RIS-assisted MIMO D2D networks, where the RIS  assists the  interference alignment 
to alleviate the co-channel interference  among multiple transceiver pairs.
We  exploited  an RIS  yielding compensation for direct-link rank-deficient channels to improve the feasibility of interference alignment conditions, thereby increasing the achievable DoFs. 
Specifically, we  presented a rank minimization  to maximize  DoFs  by 
designing the phase-shift matrix at RIS, transceiver matrix while satisfying  interference alignment conditions. 
To achieve the algorithmic advantages and admirable performance, 
a  block-structured Riemannian pursuit algorithm was developed. This is achieved by  performing a sequence of non-convex least square problems with  rank increase, followed by  unified RCG algorithms to alternately design  one block of the  fixed-rank transceiver matrix and  the unit modulus constrained phase shifters.
Simulation results showed the effectiveness of deploying an RIS and the superiority of the proposed block-structured Riemannian pursuit in terms of the achievable DoFs and sum rate. 

\bibliographystyle{IEEEtran}
\bibliography{ref} 

\end{document}